\documentclass[10pt,twocolumn]{article}
\usepackage[T1]{fontenc}
\usepackage[utf8]{inputenc}
\usepackage{lmodern}
\usepackage[a4paper,top=17mm,bottom=19mm,left=16.5mm,right=16.5mm,columnsep=7mm]{geometry}
\usepackage{microtype}
\usepackage{amsmath,amssymb}
\usepackage{booktabs}
\usepackage{enumitem}
\usepackage{xurl}
\usepackage[hidelinks]{hyperref}
\usepackage{cite}
\usepackage{listings}
\usepackage{balance}
\usepackage{etoolbox}

\hypersetup{
  pdftitle={From Agent Output to Authorized Transition},
  pdfauthor={Christopher Koch},
  pdfsubject={Agentic software engineering assurance},
  pdfkeywords={agentic software engineering, assurance, evidence admission, Agile-V, firmware, PCB engineering}
}

\setlist{nosep,leftmargin=*}
\providecommand{\tightlist}{%
  \setlength{\itemsep}{0pt}\setlength{\parskip}{0pt}}
\providecommand{\passthrough}[1]{#1}
\AtBeginEnvironment{thebibliography}{\footnotesize\interlinepenalty=10000}
\BeforeBeginEnvironment{quote}{\begin{minipage}{\columnwidth}}
\AfterEndEnvironment{quote}{\end{minipage}}
\makeatletter
\begin{document}
\twocolumn[
\begin{@twocolumnfalse}
\begin{center}
{\LARGE\bfseries From Agent Output to Authorized Transition\par}
\vspace{0.35em}
{\large\bfseries The Agile-V Assurance Spine for State-Bound Evidence in Software, Firmware, and Hardware Engineering\par}
\vspace{0.9em}
{\normalsize Christopher Koch\par}
\end{center}
\vspace{0.4em}
\begin{abstract}
Agentic engineering systems can edit repositories, run tools and tests, build firmware, synthesize schematics, and prepare deployable or manufacturable artifacts. The assurance problem is therefore shifting from whether an agent can produce an output to whether an engineering lifecycle is justified in acting on claims about that output. Current products and standards provide sandboxes, approvals, hooks, traces, policy enforcement, attestations, bills of materials, and assurance representations, but these capabilities remain fragmented. This paper presents the Agile-V Assurance Spine, a cross-domain transition contract for software, firmware, and PCB engineering. Evidence is admitted only when it establishes required properties through an authoritative source profile, is bound to the exact artifact and frozen policy baseline, remains current with respect to declared dependencies, and satisfies risk-appropriate independence and authority. Gate decisions are recorded as receipts; approvals and exceptions are exact-scope and time-bounded; and authorization is rechecked at the effect boundary before merge, deployment, flashing, release, or fabrication. A bounded review of contemporary research, commercial platforms, open-source infrastructure, and standards positions the model relative to evidence-gated lifecycle control, continuous assurance, runtime admission, provenance, and AI/ML inventories. The paper contributes a precise vocabulary, compositional architecture, domain profiles, mapping to open-source implementations, and an adversarial evaluation agenda. It does not claim regulatory conformity or demonstrated production superiority.
\end{abstract}
\vspace{1.2em}
\end{@twocolumnfalse}
]
\makeatother
\section{Introduction}\label{introduction}

Coding agents are becoming execution systems rather than suggestion systems. OpenHands, Codex, Claude Code, GitHub Copilot, Cursor, Windsurf, Jules, Devin, Cline, Aider, and related tools can perform multi-step repository work with varying combinations of sandboxing, human approval, lifecycle hooks, plan review, automated tests, session history, and pull-request integration \cite{ref20,ref21,ref22,ref23,ref24,ref25,ref26,ref27,ref28,ref29}. This is a material engineering change, but it does not make a generated patch, a passing test log, or an agent-produced review equivalent to an authorized release.

Recent research sharpens the problem. Software-agent benchmarks increasingly reveal task-quality, leakage, reward-hacking, security, and long-horizon reliability limitations \cite{ref49,ref50,ref51,ref52,ref53}. A 2026 synthesis characterizes the resulting bottleneck as the difference between code production and \textbf{production-qualified change}, where review, integration, testing, security, deployment, and operational evidence remain constraining stages \cite{ref19}. A systematic survey of specification, verification, and enforcement finds the field fragmented and reports that no reviewed approach simultaneously provides soundness, scalability, semantic correctness, and task-level safety preservation \cite{ref16}.

Several independent research lines now converge on the need for evidence-gated, state-aware engineering control \cite{ref2,ref3,ref4,ref5,ref6,ref7,ref8,ref9,ref10,ref11,ref12,ref13,ref14,ref15,ref16,ref17,ref18,ref19}. This paper uses the SCOPE-V lifecycle - \textbf{Specify, Constrain, Orchestrate, Prove, Evolve, Verify} - as its immediate process substrate \cite{ref1}, but grounds the proposed assurance contract primarily in external research, standards, and current market systems.

The remaining question is more precise:

\begin{quote}
\textbf{When is evidence allowed to count toward an engineering transition, and when may that transition actually execute?}
\end{quote}

A file can be present but malformed. A schema-valid result can be irrelevant. A relevant result can be stale. A correct test can be bound to the wrong commit. A signed approval can authorize an earlier policy or artifact. A fresh verifier context can reduce confirmation bias without providing organizational independence. A runtime hook can block an operation without explaining which engineering claim was established. A trace can show what happened without proving that the outcome was acceptable. An SBOM or AI/ML-BOM can inventory components without authorizing use of the assembled system.

\subsection{Contributions}\label{contributions}

This paper makes five contributions.

\begin{enumerate}
\def\labelenumi{\arabic{enumi}.}
\tightlist
\item
  \textbf{A market- and literature-grounded gap analysis.} It compares the Assurance Spine against major coding-agent platforms, semantic change-control products, observability stacks, policy engines, provenance and attestation systems, AI/ML-BOM standards, agent protocols, assurance-case standards, and the closest contemporary research.
\item
  \textbf{A property-level evidence-admission model.} It separates presence, structural completeness, sufficiency, admissibility, gate eligibility, and transition authorization.
\item
  \textbf{A trusted transition contract.} It binds evidence, policy, approvals, exceptions, and authority to an exact subject state and frozen verification baseline, with risk-aligned independence and claim-scoped revalidation.
\item
  \textbf{A compositional architecture.} It shows how existing tools - rather than being replaced - can supply execution, traces, policies, attestations, inventories, and protocol messages to the Assurance Spine.
\item
  \textbf{Cross-domain profiles and an evaluation agenda.} It applies one vocabulary to software, firmware, and PCB engineering and defines adversarial conformance and empirical evaluation work.
\end{enumerate}

\subsection{Claim boundary}\label{claim-boundary}

The Assurance Spine does not prove universal software correctness, eliminate trust bootstrap, make an AI reviewer organizationally independent, or establish compliance merely because schemas validate. It is a design and executable-reference contribution intended to make transition decisions explicit, falsifiable, and state-bound. Current repository tests support internal contract conformance for encoded scenarios; they are not evidence of production safety or cross-runtime equivalence.

\subsection{Artifact availability}\label{artifact-availability}

The reference contracts and execution scaffold are available in the public \passthrough{\lstinline!agile\_v\_skills!} and \passthrough{\lstinline!agentic\_agile\_v!} repositories at the immutable snapshots identified in \cite{ref55}. The repositories are research artifacts, not certified products, and the paper distinguishes repository-level conformance from live runtime validation.

\section{Landscape Review and Positioning}\label{landscape-review-and-positioning}

\subsection{Review method}\label{review-method}

The comparison is a bounded market and research scan, not a systematic review or a claim to have enumerated every private feature or unpublished system. The scan combined keyword searches for agentic software engineering, evidence admission, continuous assurance, runtime governance, agent security, software provenance, AI/ML bills of materials, firmware verification, and PCB synthesis with backward and forward citation chaining from the closest research \cite{ref1,ref2,ref3,ref4,ref5,ref6,ref7,ref8,ref9,ref10,ref11,ref12,ref13,ref14,ref15,ref16,ref17,ref18,ref19,ref49,ref50,ref51,ref52,ref53,ref54}. Product claims were checked against vendor documentation, standards against official specifications, and repository claims against public source snapshots. Inclusion required direct relevance to transition control or candidate evidence. The review closed on 23 September 2026.

The reviewed landscape includes:

\begin{itemize}
\tightlist
\item
  ten representative coding-agent and agent-development platforms \cite{ref20,ref21,ref22,ref23,ref24,ref25,ref26,ref27,ref28};
\item
  six representative AI-code quality, security, review, and governance products \cite{ref56,ref57,ref58,ref59,ref60,ref61};
\item
  semantic change-control and operational-memory infrastructure \cite{ref29};
\item
  agent observability and evaluation systems \cite{ref30,ref31,ref32};
\item
  policy/admission and software-supply-chain mechanisms \cite{ref33,ref34,ref35,ref36};
\item
  AI/ML inventory standards and runtime collectors \cite{ref37,ref38,ref39};
\item
  interoperability protocols \cite{ref40,ref41,ref42};
\item
  control, provenance, and assurance-case standards \cite{ref43,ref44,ref45,ref46};
\item
  security/governance frameworks and regulation \cite{ref47,ref48}; and
\item
  the closest scholarly work on evidence sufficiency, lifecycle gating, runtime admission, continuous assurance, and agent security \cite{ref2,ref3,ref4,ref5,ref6,ref7,ref8,ref9,ref10,ref11,ref12,ref13,ref14,ref15,ref16,ref17,ref18,ref19,ref49,ref50,ref51,ref52,ref53,ref54}.
\end{itemize}

The comparison asks which layer each system primarily solves and whether its public specification defines the full transition relation proposed here.

\subsection{Execution agents: strong control points, limited shared assurance semantics}\label{execution-agents-strong-control-points-limited-shared-assurance-semantics}

Current coding-agent products already expose important enforcement surfaces.

\begin{itemize}
\tightlist
\item
  \textbf{OpenHands} supports repository hooks that can block tool use, deny completion, require tests or linting, and log activity \cite{ref20}.
\item
  \textbf{Codex} documents sandboxing, permission profiles, approvals, hooks, auto-review, and controlled internet access \cite{ref21}.
\item
  \textbf{Claude Code} combines read-only defaults or classifier-mediated autonomy with sandboxing, working-directory boundaries, managed permissions, cloud isolation, audit logging, and hooks \cite{ref22}.
\item
  \textbf{GitHub Copilot} provides repository hooks that can approve or deny tool execution, invoke secret scanning, enforce validation rules, and create audit logs \cite{ref23}.
\item
  \textbf{Cursor} and \textbf{Windsurf} expose lifecycle hooks around tools, commands, file access, prompts, and stop events \cite{ref24,ref25}.
\item
  \textbf{Jules} provides explicit plan review and approval before implementation \cite{ref26}.
\item
  \textbf{Devin} provides isolated sessions, parallel work, reusable playbooks, knowledge, schedules, and event timelines \cite{ref27}.
\item
  \textbf{Cline} exposes per-tool and conditional approval policies \cite{ref28}, while \textbf{Aider} tightly couples edits to Git history and attribution.
\end{itemize}

These are valuable implementation mechanisms. Their public documentation mainly defines \textbf{how to execute, constrain, observe, or approve agent actions}, not a portable engineering semantics for which claims a given test, trace, review, or attestation is allowed to establish. The Assurance Spine can use these products as execution backends while keeping the evidence and gate contract outside any one vendor.

The quality and governance market increasingly adds controls around those agents. \textbf{GitLab AI Governance} records agent sessions and audit events and enforces \passthrough{\lstinline!Allow!}, \passthrough{\lstinline!Ask!}, and \passthrough{\lstinline!Deny!} tool policies across GitLab Duo and external agents connected through GitLab MCP \cite{ref56}. \textbf{SonarQube AI Code Assurance} marks AI-bearing projects and applies AI-qualified quality gates for security, reliability, coverage, hotspot review, and duplication \cite{ref57}. \textbf{Qodo} combines agentic pull-request review with organizational rule and compliance checks \cite{ref58}. \textbf{Semgrep Guardian} integrates MCP, hooks, and skills to scan every file write and block vulnerable patterns, secrets, and malicious dependencies \cite{ref59}. \textbf{Snyk Studio} exposes security analysis to coding agents through MCP and records agentic scan usage \cite{ref60}, while \textbf{Tabnine Agent} supports autonomous development work, approval checkpoints, policy validation, and enterprise-controlled context \cite{ref61}. These products materially strengthen code-quality, security, and tool-governance enforcement. Their documented gates are nevertheless product- or property-specific; they do not expose a shared contract that determines which source may establish which property, binds that evidence to a frozen cross-domain baseline, and authorizes merge, deployment, flashing, or fabrication under one transition model.

\subsection{Semantic state, memory, and change control}\label{semantic-state-memory-and-change-control}

Nool is the closest market comparator at the operational-state layer. Its public architecture centers on a semantic DAG binding intent, code, attribution, policy validation, causal lineage, evidence, and semantic blast radius, with Git interoperability and WASM-based gates \cite{ref29}. It also treats intent, impact, findings, and evidence as durable repository memory with scope, supersession, and time-aware retrieval.

This directly overlaps with Agile-V's need for persistent intent, causal history, and impact analysis. The distinction is one of emphasis and specification scope. Nool's public material presents a semantic engineering ledger and fleet/change-control product. The Assurance Spine specifies a domain-general \textbf{evidence-admission and transition-authorization relation}, including authoritative evidence-source capability profiles, risk-based independence classes, exact approval and exception semantics, revalidation dependencies, AI-influence inventory, and profiles for firmware and PCB effects. The two approaches are complementary; a Nool-style state kernel could supply subject history, intent, and blast-radius data to the Spine.

\subsection{Observability and evaluation}\label{observability-and-evaluation}

LangSmith, Phoenix, and OpenTelemetry-based tooling provide traces, spans, sessions, datasets, annotations, online evaluations, and production metrics \cite{ref30,ref31,ref32}. These systems are essential because an assurance process cannot evaluate invisible execution.

However, observation is not admission. A trace can be authentic and still fail to establish a required claim. An evaluator score can be useful but model-dependent. A successful tool span does not prove that the current requirement baseline was satisfied, that the caller possessed the needed authority, or that a later effect matched the reviewed payload. The Assurance Spine treats traces and evaluation outputs as \textbf{candidate evidence} whose admissible properties are constrained by a trusted source profile.

\subsection{Policy engines and infrastructure admission}\label{policy-engines-and-infrastructure-admission}

OPA/Gatekeeper and related Kubernetes policy engines demonstrate mature, deterministic admission patterns: evaluate an incoming object against policy at a specific boundary, then allow, deny, warn, mutate, or audit \cite{ref33}. Kyverno and Sigstore-based admission extend this with image verification and attestations. These systems validate the importance of fail-closed enforcement at the effect boundary.

Their scope is narrower than the proposed engineering contract. A Kubernetes admission controller can determine whether a resource satisfies cluster policy, but it does not normally reconstruct a software task's requirement baseline, property-level evidence sufficiency, verifier independence, human approval scope, exception coverage, or firmware/PCB verification lineage. The Assurance Spine generalizes the admission concept while allowing OPA, Gatekeeper, or Kyverno to implement specific effect-boundary checks.

\subsection{Provenance, attestations, and BOMs}\label{provenance-attestations-and-boms}

SLSA provenance records how a software artifact was produced, including the subject, builder, build definition, parameters, and dependencies \cite{ref34}. in-toto provides a framework for verifiable claims about software-supply-chain steps \cite{ref35}, and Sigstore/GitHub attestations provide signed provenance and transparency mechanisms \cite{ref36}. These are strong sources for artifact identity, integrity, and production provenance.

CycloneDX ML-BOM and SPDX AI profiles extend inventory toward models, datasets, frameworks, configurations, and provenance \cite{ref37,ref38}. \passthrough{\lstinline!k8s-aibom!} generates runtime CycloneDX ML-BOMs for inference services, agent stacks, RAG infrastructure, training jobs, and evaluation harnesses, distinguishing \passthrough{\lstinline!declared!}, \passthrough{\lstinline!inferred!}, and \passthrough{\lstinline!unresolved!} attributes and attaching evidence locators \cite{ref39}.

The Assurance Spine does not replace any of them. It consumes them. A signed SLSA attestation may establish build provenance; an in-toto statement may establish a supply-chain step; an ML-BOM may establish an observed model/runtime inventory. None should automatically be treated as proof of requirement satisfaction, human authority, or system safety beyond the properties its trusted profile permits.

\subsection{Protocols and assurance standards}\label{protocols-and-assurance-standards}

MCP standardizes resources, prompts, tools, tasks, and communication between hosts, clients, and servers, while explicitly leaving robust authorization and many security controls to implementers \cite{ref40}. A2A standardizes inter-agent tasks, messages, artifacts, status, and security metadata \cite{ref41}. AG-UI standardizes agent-user interaction events \cite{ref42}. These protocols make execution interoperable; they do not, by themselves, define which evidence is sufficient to authorize a lifecycle or physical-world transition.

OSCAL provides machine-readable security controls and assessment artifacts \cite{ref43}. W3C PROV-O models entities, activities, and agents \cite{ref44}. SACM and GSN structure assurance arguments and evidence \cite{ref45,ref46}. These standards supply useful representations, but an operational runtime still needs deterministic rules that decide whether the evidence currently available is admissible for a concrete transition.

\subsection{Closest contemporary research}\label{closest-contemporary-research}

The research landscape has moved rapidly toward the same boundary.

\begin{itemize}
\tightlist
\item
  \textbf{Proof-or-Stop} treats lifecycle states as claims until current, tracked-state-bound evidence permits transition and reports adversarial receipt-tampering results \cite{ref2}.
\item
  \textbf{DEMM} and \textbf{DEMM-Bench} formalize property-level evidence sufficiency and show why trace-, schema-, or ledger-presence baselines can overclaim \cite{ref3,ref4}.
\item
  \textbf{Cognitive Admission Control (CAC)} is the closest admission-calculus comparator. It maps typed actions and risk to predicates, evidence classes, scope, freshness, and witness constraints; successful admission yields a certificate and dispatch guards \cite{ref13}.
\item
  \textbf{Agent-Integrated Software and Intent-Level Interaction Abstraction} bind effects to task revision, referenced objects, role authority, controller identity, and outcome evidence, and separate admission from occurrence \cite{ref14}.
\item
  \textbf{TAIP} frames continuous AI assurance as reusable assurance objects and claim/evidence posture across organizational levels \cite{ref15}.
\item
  \textbf{AI-GRACE} maps organizational objectives and obligations through risk, assurance, controls, evidence, capabilities, and architecture \cite{ref5}.
\item
  \textbf{Runtime Governance: Policies on Paths} treats the partial execution path and proposed next action as central policy inputs \cite{ref17}.
\item
  \textbf{Verifiable Manifest Signing for MCP} binds tool-use manifests to freshness, policy, signatures, fail-closed verification, and transparency logs \cite{ref18}.
\item
  \textbf{Safe-agent surveys and security benchmarks} expose the limits of prompt-only controls, specification translation, and local test success \cite{ref16,ref49,ref50,ref51,ref52,ref53}.
\end{itemize}

These works narrow the novelty claim. The Assurance Spine is not the first proposal for evidence-gated control, continuous assurance, admission certificates, state binding, or policy-mediated runtime actions. Its proposed contribution is their \textbf{engineering-lifecycle synthesis}: a portable transition contract that combines property-level evidence admission, trusted source capabilities, exact state and policy binding, independence and authenticated authority, explicit exceptions, dependency-scoped revalidation, AI-influence inventory, and cross-domain effect profiles in one open reference model.

\subsection{Market gap statement}\label{market-gap-statement}

Based on the reviewed public documentation and literature, the market contains strong components for nearly every part of the problem. We did not find a reviewed public system that documents all of the following as one cross-platform, cross-domain contract:

\begin{lstlisting}
required claim properties
+ authoritative source capability profiles
+ exact subject and policy binding
+ contradiction-aware admission
+ risk-aligned independence / authority
+ explicit exception semantics
+ dependency-scoped evidence revalidation
+ model/runtime/tool influence inventory
+ atomic recheck at the effect boundary
\end{lstlisting}

The Assurance Spine is intended as the composition layer between those capabilities, not as a replacement product for them.

\section{The Assurance Spine Model}\label{the-assurance-spine-model}

\subsection{Admission hierarchy}\label{admission-hierarchy}

The fundamental hierarchy is:

\begin{lstlisting}
present
  -> structurally complete
     -> sufficient for the required properties
        -> admissible for state + policy
           -> eligible for this gate
              -> effect-authorized
\end{lstlisting}

Only the forward direction is safe. Presence does not imply completeness; completeness does not imply sufficiency; sufficiency does not imply current admissibility; admissibility of one item does not imply gate eligibility; and eligibility does not prove that the authorized effect actually occurred.

\subsection{Core objects}\label{core-objects}

For task \(\tau\), define an exact \textbf{subject state}:

\[
\sigma = (kind, ref, digest)
\]

The subject can be a Git commit, firmware image, PCB manufacturing package, document revision, dataset, container image, or other controlled artifact.

A frozen \textbf{verification baseline} is:

\[
B = (R, C, \rho, \pi, P)
\]

where \(R\) is the requirement baseline, \(C\) the acceptance criteria, \(\rho\) the risk state, \(\pi\) the policy/control digest, and \(P\) the applicable evidence-property profiles. Once Prove begins, the active baseline is frozen; Evolve may propose future changes but cannot weaken the criteria being verified in the current cycle.

A claim \(c\) has a required property set \(Req(c)\). An evidence item \(e\) declares:

\begin{lstlisting}
supports(e)        claims it addresses
result(e)       pass | fail | error | unknown
properties(e)   properties it claims
source(e)       adapter ID + profile digest
state(e)           subject binding
policy(e)          policy binding
integrity(e)    digest / signature
                or attestation
independence(e)    achieved independence class
invalidators(e) source / requirement /
                policy / environment / tool /
                model / hardware
\end{lstlisting}

\subsection{Evidence sources cannot self-authorize their meaning}\label{evidence-sources-cannot-self-authorize-their-meaning}

Let \(Cap(a)\) be the independently resolved capability profile for evidence adapter \(a\). For a passing evidence item:

\[
TrustedProps(e) = properties(e) \cap Cap(source(e))
\]

In the strict runtime contract, an item that claims a property outside the source profile is rejected rather than silently upgraded or trusted. This blocks a common category error: a unit-test adapter cannot establish \passthrough{\lstinline!human\_authority!}; a static analyzer cannot establish \passthrough{\lstinline!intended\_use\_validated!}; an LLM review cannot establish \passthrough{\lstinline!organizational\_independence!}; a BOM cannot establish \passthrough{\lstinline!requirement\_satisfied!} merely by listing components.

For mandatory claim \(c\), evidence is property-sufficient only when:

\[
Req(c) \subseteq \bigcup_{e:\,c\in supports(e),\ result(e)=pass} TrustedProps(e)
\]

and no supporting evidence reports a contradictory \passthrough{\lstinline!fail!} or \passthrough{\lstinline!error!}. This avoids the \textbf{any-pass anti-pattern}, where one positive result masks a negative result for the same claim.

\subsection{Admissibility}\label{admissibility}

Property sufficiency is necessary but not enough. A claim is admissible only if its supporting evidence also satisfies:

\begin{enumerate}
\def\labelenumi{\arabic{enumi}.}
\tightlist
\item
  exact subject type, reference, and digest match;
\item
  exact frozen policy digest match;
\item
  integrity and producer attribution requirements;
\item
  freshness and dependency-coverage requirements;
\item
  required independence floor;
\item
  applicable environment and intended-use bindings; and
\item
  trusted-resolution rules for source and property profiles.
\end{enumerate}

Unknown or partial dependency coverage is not evidence of unchanged state. For higher-risk work, incomplete coverage widens revalidation rather than silently permitting reuse.

\subsection{Gate Receipts and trusted decision context}\label{gate-receipts-and-trusted-decision-context}

A \textbf{Gate Receipt} records why a transition was allowed, denied, waived, marked stale, or escalated. It binds:

\begin{lstlisting}
task and gate
exact subject state
requirement baseline
policy digest
risk level
required / admitted / rejected / stale claims
open obligations
verifier independence
approval references
exception references
decision and reason codes
\end{lstlisting}

The receipt is evaluated against a \textbf{Trusted Admission Context} supplied outside candidate write authority. This context contains the accepted task, gate, subject, policy, risk, action, resources, critical risks, and non-waivable controls. Resolvers for approvals, exceptions, source profiles, and property profiles must be authoritative and fail closed.

Eligibility can be summarized as:

\[
Eligible(g)=C_a \land I_s \land A_v \land R_d \land X_v
\]

where the terms denote claims admitted, independence sufficient, authority valid, risks dispositioned, and exceptions valid.

\subsection{Eligibility is not execution}\label{eligibility-is-not-execution}

The runtime must distinguish:

\begin{lstlisting}
eligible to execute
executed
outcome observed
outcome acceptable
\end{lstlisting}

An approval can be authentic yet stale. A dispatch can be admitted but fail before the external effect. A timeout can leave outcome unknown. Therefore, immediately before merge, deployment, flashing, fabrication submission, or other consequential effect, the runtime should atomically recheck:

\begin{lstlisting}
subject digest
policy digest
authority and revocation state
single-use approval consumption
action and resource scope
dispatch guard
\end{lstlisting}

The resulting effect record should bind the same subject and action as the Gate Receipt. This aligns the Assurance Spine with CAC's dispatch guards \cite{ref13} and interaction-effect obligations in Agent-Integrated Software \cite{ref14}.

\subsection{Threat model and trust boundary}\label{threat-model-and-trust-boundary}

The candidate-controlled surface includes agent outputs, evidence claims, adapter assertions, task-local policy references, and references to approvals or exceptions. The design assumes that these inputs may be incomplete, stale, replayed, contradictory, or intentionally misleading. The trusted computing base consists of versioned source/property-profile registries, digest and clock services, authenticated authority providers, append-only decision records, and the effect mediator that performs the final atomic recheck.

The principal security goals are: \textbf{no source self-elevation}, \textbf{no state or policy replay}, \textbf{no hidden waiver}, \textbf{no approval reuse outside scope}, and \textbf{no effect substitution after review}. Unresolved mandatory claims block advancement. Out of scope are incorrect requirements accepted into the baseline, compromise or collusion of all trusted providers, fraudulent decisions by an otherwise authorized human, and physical deception that is outside the modeled evidence sources. Those cases require additional organizational, cryptographic, or domain controls rather than stronger agent prompting.

\section{Governance Semantics}\label{governance-semantics}

\subsection{Risk-aligned independence}\label{risk-aligned-independence}

The Spine distinguishes five assurance classes:

\begin{itemize}
\tightlist
\item
  \textbf{I0 - self-check:} same agent and context; useful sanity check, not independence.
\item
  \textbf{I1 - context-separated:} fresh context; reduces memory and confirmation bias.
\item
  \textbf{I2 - role-separated:} separate verifier role with protected requirement/test inputs.
\item
  \textbf{I3 - authority-separated:} authenticated principal who cannot modify the evaluated artifact, policy, or criteria.
\item
  \textbf{I4 - organizationally independent:} separate qualified human, team, or entity where a governing profile requires it.
\end{itemize}

A second AI call does not become I3 or I4 merely by using a different prompt or model. Risk and domain profiles set the minimum independence for each critical claim, not once for the whole task.

\subsection{Exceptions remain visible}\label{exceptions-remain-visible}

Waiver, concession, dispensation, residual-risk acceptance, and defer are different decisions:

\begin{itemize}
\tightlist
\item
  a \textbf{waiver} suspends a criterion for bounded scope and time;
\item
  a \textbf{concession} accepts a known nonconformity for a specific case;
\item
  a \textbf{dispensation} changes an obligation's timing, not its substance;
\item
  \textbf{residual-risk acceptance} accepts consequence, not failed verification as truth; and
\item
  \textbf{defer} leaves work unresolved and does not authorize advancement.
\end{itemize}

Every unresolved mandatory item must be covered by an applicable, current, authenticated exception. Gate output is \passthrough{\lstinline!WAIVED!}, not \passthrough{\lstinline!PASS!}. Identity, receipt integrity, subject binding, and gate integrity are non-waivable meta-controls.

\subsection{Change-aware revalidation}\label{change-aware-revalidation}

Evidence depends on more than source code. Each item can declare invalidation dependencies of type:

\begin{lstlisting}
source | requirement | policy | environment |
tool | model | dataset | hardware | authority
\end{lstlisting}

A change assessment yields \passthrough{\lstinline!UNCHANGED!}, \passthrough{\lstinline!REVALIDATION\_REQUIRED!}, \passthrough{\lstinline!STALE!}, or \passthrough{\lstinline!UNKNOWN!}. Only unchanged dependencies with complete coverage permit evidence reuse. This supports targeted revalidation: a compiler change may invalidate firmware build and HIL evidence without invalidating a requirements approval; a PCB revision may invalidate firmware pin-map tests; a model/runtime change may invalidate AI-assisted review evidence but not a cryptographically verified build attestation.

\subsection{AI-influence inventory}\label{ai-influence-inventory}

SBOMs describe software components; AI-assisted engineering also depends on models, model deployments, agent runtimes, skills, tools, retrieval sources, sandboxes, evaluators, and external APIs. The Spine therefore links an \textbf{AI Run Manifest / Agent Run BOM} to the evidence bundle. It records auditable metadata, not hidden chain-of-thought:

\begin{lstlisting}
model and provider identity
runtime and deployment version
loaded skills and policies
tool and MCP server versions
retrieval / dataset references
sandbox or container identity
evaluation configuration
influenced artifacts
confidence: declared | inferred | verified |
            unresolved
evidence locator and digest
\end{lstlisting}

CycloneDX ML-BOM, SPDX AI, and \passthrough{\lstinline!k8s-aibom!} provide interoperable inventory foundations \cite{ref37,ref38,ref39}. The additional Agile-V function is to connect inventory changes to affected claims and revalidation scope.

\subsection{Governance conversion without same-cycle goalpost changes}\label{governance-conversion-without-same-cycle-goalpost-changes}

Controls can originate \textbf{ex ante} from organizational obligations, architecture, and known risk \cite{ref5,ref10,ref17}, or \textbf{ex post} from incidents, verifier findings, and recurring agent failure patterns \cite{ref11}. Both enter the same governed process:

\begin{lstlisting}
observation or obligation
  -> proposed control
     -> authority review
        -> held-out validation
           -> versioned activation
              -> future baseline
\end{lstlisting}

Learning during Evolve is encouraged. Changing the active criteria after a failed Prove step is not. This prevents an agent or team from converting a current failure into a same-cycle rule relaxation.

\section{Compositional Architecture}\label{compositional-architecture}

The Assurance Spine is a control plane over existing capabilities:

\begin{lstlisting}
Human / organizational authority
        |
Requirements, risk, policy, evidence profiles
        |
Agile-V Assurance Spine
  - baseline freeze
  - evidence admission
  - revalidation
  - gate receipt
  - effect authorization
        |
+-------+---------+-----------+-----------+
|                 |           |           |
Execution      Observability   Policy
agents         and evals       engines
              Provenance / assurance
        |
Effects: merge | deploy | flash | fabricate |
         release
\end{lstlisting}

\subsection{Interoperability mapping}\label{interoperability-mapping}

\begin{itemize}
\tightlist
\item
  \textbf{OpenHands, Codex, Claude Code, Copilot, Cursor, Windsurf, Jules, Devin, and Cline} execute work and expose approval/hook boundaries \cite{ref20,ref21,ref22,ref23,ref24,ref25,ref26,ref27,ref28}.
\item
  \textbf{Nool} can supply durable intent, impact, lineage, memory, and semantic change state \cite{ref29}.
\item
  \textbf{LangSmith, Phoenix, and OpenTelemetry} supply traces, evaluations, sessions, and metrics \cite{ref30,ref31,ref32}.
\item
  \textbf{OPA/Gatekeeper, Kyverno, and policy-as-code systems} enforce deterministic controls at selected boundaries \cite{ref8,ref10,ref17,ref33}.
\item
  \textbf{SLSA, in-toto, Sigstore, and GitHub attestations} supply artifact and process provenance \cite{ref34,ref35,ref36}.
\item
  \textbf{CycloneDX, SPDX, and \passthrough{\lstinline!k8s-aibom!}} supply software and AI/ML inventories \cite{ref37,ref38,ref39}.
\item
  \textbf{MCP, A2A, and AG-UI} transport tools, tasks, artifacts, agent messages, and user interaction events \cite{ref40,ref41,ref42}.
\item
  \textbf{OSCAL, PROV-O, SACM, and GSN} can encode controls, provenance, claims, argument structures, and evidence references \cite{ref43,ref44,ref45,ref46}.
\end{itemize}

The Spine's distinct role is to answer: \textbf{may these records count, for this claim, now, for this exact transition?}

\subsection{Repository realization}\label{repository-realization}

The current open-source Agile-V repositories divide the reference implementation between contract semantics and execution infrastructure. \passthrough{\lstinline!agile\_v\_skills!} contains Evidence Bundle v2, source/property profiles, Gate Receipts, Approval v2, independence classes, exceptions, revalidation, risk floors, governance conversion, Trusted Admission Context, AI-influence manifests, and aggregate admission evaluators; \passthrough{\lstinline!agentic\_agile\_v!} provides task briefs, evidence workflows, OpenHands hooks, graph/impact infrastructure, OpenWiki integration, and PCB development components \cite{ref55}.

At the snapshots reviewed for this paper, the skills repository contains a more advanced admission contract than the execution repository. The next implementation priority is therefore not another prompt layer: it is faithful porting of the aggregate evaluators and trusted-provider boundaries into the runtime, followed by effect-boundary adapters and cross-runtime conformance tests.

\section{Software, Firmware, and PCB Profiles}\label{software-firmware-and-pcb-profiles}

The same core contract applies across domains; the evidence and effect profiles differ.

\subsection{Software}\label{software}

\textbf{Subject:} source commit, build artifact, container image, deployment manifest.\\
\textbf{Evidence:} tests, static analysis, security scans, review, build provenance, runtime checks.\\
\textbf{Effect:} merge, package publication, deployment, migration.\\
\textbf{Critical bindings:} commit/build digest, requirement baseline, policy, environment, approval scope, rollback plan.

Coding-agent hooks can collect and block on evidence, but release authorization should depend on aggregate admission rather than \passthrough{\lstinline!tests passed!} alone. SWE-Bench Pro Verified, AgentDojo, ToolEmu, AgentSecBench, and CIBER offer useful capability and adversarial scenarios, yet production admission also needs organization-specific requirements and authority \cite{ref49,ref50,ref51,ref52,ref53}.

\subsection{Firmware}\label{firmware}

\textbf{Subject:} source state, board revision, toolchain/SDK, configuration, linked binary, signed image.\\
\textbf{Evidence:} host tests, cross-build, static analysis, simulation, timing traces, hardware-in-the-loop tests, boot/update/rollback evidence.\\
\textbf{Effect:} image signing, release, device flashing, over-the-air update.\\
\textbf{Critical bindings:} hardware-firmware contract, pin/peripheral map, memory layout, compiler flags, binary digest, device class, HIL rig identity.

Firmware shows why a successful build is insufficient: code can compile against the wrong board revision, bus mapping, interrupt polarity, ADC scaling, or memory partition.

\subsection{PCB engineering}\label{pcb-engineering}

\textbf{Subject:} Circuit IR, KiCad schematic/board revision, component set, manufacturing package digest.\\
\textbf{Evidence:} schema checks, ERC/DRC, semantic electrical review, power/thermal analysis, footprint and component provenance, fabrication outputs, independent EE review.\\
\textbf{Effect:} design release, purchase, fabrication submission, assembly release.\\
\textbf{Critical bindings:} netlist, BOM, datasheets, footprints, stack-up, Gerber/drill archive, risk class, manufacturing approval.

pcbGPT demonstrates the value of a structured intermediate representation, component/datasheet grounding, deterministic execution checks, semantic validation, and generate-execute-repair loops for reviewable KiCad drafts \cite{ref54}. The Assurance Spine adds transition semantics: ERC or an LLM semantic review becomes typed evidence, while fabrication remains blocked until the exact manufacturing package is admitted and authorized by the required engineering authority.

\subsection{Cross-domain co-verification}\label{cross-domain-co-verification}

For embedded products, a single requirement can span all three domains:

\begin{lstlisting}
requirement
  -> PCB net / component
  -> board contract / pin map
  -> firmware driver and configuration
  -> protocol/API exposed to software
  -> system test / HIL evidence
\end{lstlisting}

A change to any linked state should invalidate only the claims that depend on it, but unknown coverage must widen verification. This is the practical value of one Assurance Spine across domains.

\subsection{Illustrative vertical slice: adding an I2C sensor}\label{illustrative-vertical-slice-adding-an-i2c-sensor}

Consider a requirement to add a temperature sensor to an embedded product and expose readings through a software API. This example is illustrative rather than empirical.

\begin{enumerate}
\def\labelenumi{\arabic{enumi}.}
\tightlist
\item
  The PCB profile binds the selected component, address, supply rail, pull-ups, interrupt net, footprint, and manufacturing package to a board-revision digest. ERC/DRC, component provenance, and an independent electrical review establish only the properties permitted by their source profiles.
\item
  The firmware profile imports the approved pin and bus contract, builds a driver against the exact board revision, and binds tests, simulation, HIL logs, and the signed image to the binary digest and toolchain state.
\item
  The software profile binds the API schema, serialization behavior, integration tests, and deployment artifact to the corresponding firmware protocol revision.
\item
  Separate Gate Receipts authorize fabrication, firmware release, and software deployment. A receipt for one effect cannot authorize another.
\item
  If the board revision changes the interrupt pin, only evidence depending on that hardware state becomes stale. Firmware pin-map and HIL claims must be reproduced; an unchanged API contract or previously approved requirement may remain eligible. If dependency coverage is incomplete, the affected scope widens conservatively.
\end{enumerate}

The example shows the intended composition: each domain uses its own evidence methods, while claim properties, state binding, revalidation, authority, and effect integrity follow one transition contract.

\section{Evaluation Plan}\label{evaluation-plan}

\subsection{What is already demonstrated}\label{what-is-already-demonstrated}

The reference implementation described in Section 5.2 provides schemas, semantic predicates, aggregate admission functions, golden journeys, negative journeys, and contract tests. This demonstrates that the proposed semantics can be encoded and deterministically tested. It does not demonstrate live conformance in OpenHands, Codex, Claude Code, Copilot, Nool, or other products; nor does it demonstrate lower defect rates in production.

\subsection{Adversarial admission benchmark}\label{adversarial-admission-benchmark}

A meaningful benchmark should compare the full aggregate decision against weaker baselines:

\begin{lstlisting}
B0 agent says DONE
B1 trace or evidence file exists
B2 schema validates
B3 evidence is signed
B4 policy hook passes
B5 reviewer artifact exists
B6 Assurance Spine aggregate admission
\end{lstlisting}

Test cases should include:

\begin{itemize}
\tightlist
\item
  correct evidence bound to the wrong artifact digest;
\item
  stale policy or requirement baseline;
\item
  a test adapter claiming human authority;
\item
  one passing result masking a contradictory failure;
\item
  replayed, expired, consumed, or wrong-scope approval;
\item
  fresh-context verification mislabeled as organizational independence;
\item
  incomplete dependency coverage reported as unchanged;
\item
  a defer or residual-risk acceptance misused as a verification waiver;
\item
  omitted model/runtime/tool changes;
\item
  a race between approval and effect execution;
\item
  correct software evidence paired with the wrong PCB or firmware revision; and
\item
  fabricated provenance, authority, or adapter identity.
\end{itemize}

Proof-or-Stop, DEMM-Bench, CAC, AgentDojo, ToolEmu, AgentSecBench, and SWE-Bench Pro Verified offer complementary failure models and benchmark design patterns \cite{ref2,ref4,ref13,ref49,ref50,ref51,ref52,ref53}.

\subsection{Measures}\label{measures}

Evaluation should report:

\begin{itemize}
\tightlist
\item
  false transition acceptance and false rejection;
\item
  stale-evidence and approval-replay detection;
\item
  precision of claim-scoped revalidation;
\item
  unsupported property-claim detection;
\item
  authorization latency and runtime overhead;
\item
  human reviewer time and escalation load;
\item
  evidence collection and storage cost;
\item
  task completion under fixed risk and budget;
\item
  cross-runtime semantic agreement; and
\item
  exact-effect integrity: whether the reviewed digest is the one merged, deployed, flashed, or fabricated.
\end{itemize}

\subsection{Empirical program}\label{empirical-program}

Four stages are proposed:

\begin{enumerate}
\def\labelenumi{\arabic{enumi}.}
\tightlist
\item
  \textbf{Reference conformance:} run all golden and negative journeys against the repository semantics.
\item
  \textbf{Cross-runtime conformance:} implement adapters for at least OpenHands and one additional hook-capable coding agent, then compare identical decisions.
\item
  \textbf{Domain vertical slices:} evaluate one production-like software task, one PCB-to-firmware change, and one signed firmware update path.
\item
  \textbf{Field study:} compare ordinary CI/review against Assurance-Spine-assisted delivery on defect escape, cycle time, reviewer effort, and revalidation scope.
\end{enumerate}

The hypothesis is not that more gates always improve delivery. The testable proposition is that property-aware, state-bound admission reduces specific false-authority and stale-evidence failures at acceptable cost.

\section{Threats, Limitations, and Conclusion}\label{threats-limitations-and-conclusion}

The market scan is representative and based on public sources; private product capabilities may exceed their documentation. Rapid product change can make individual feature comparisons stale. Some 2026 research cited here is preprint work and should not be treated as settled evidence. Repository conformance is not production validation. Source and authority resolvers remain trust anchors; compromised providers can still produce false decisions. Semantic correctness remains bounded by the quality of requirements, evidence methods, environment models, and human judgment. The framework can support regulated engineering activities but does not itself establish legal compliance, certification, or safety.

The external comparison nevertheless changes the paper's conclusion in an important way. The ecosystem does not lack agent sandboxes, permissions, hooks, traces, policy engines, attestations, inventories, or assurance representations. It lacks a widely shared contract that composes those records into a defensible answer to one operational question:

\begin{quote}
\textbf{May this exact artifact, under this exact policy and risk context, be advanced by this authority on the basis of this evidence?}
\end{quote}

The Agile-V Assurance Spine answers that question by treating agent outputs as claims, evidence sources as capability-bounded producers, baselines as frozen state, changes as revalidation triggers, approvals and exceptions as exact-scope authority records, and lifecycle decisions as explicit receipts. It then requires a final recheck where the consequential effect occurs.

This is deliberately narrower than general AI safety and broader than a coding-agent hook. It is an engineering transition contract. Its value will depend on whether future conformance and field studies show that it catches stale, misbound, insufficient, or falsely authorized evidence without making useful delivery impractically slow. That is the next research and implementation task.

\section*{Generative AI Disclosure}
OpenAI ChatGPT was used for research assistance, source organization, drafting, language editing, and LaTeX preparation. The author assumes full responsibility for the final manuscript, including verification of claims and citations.

\balance


\begin{thebibliography}{99}
\setlength{\itemsep}{0.25em}
\setlength{\parskip}{0pt}
\bibitem{ref1}
C. Koch, ``Agentic Agile-V: From Vibe Coding to Verified Engineering in Software and Hardware Development,'' arXiv:2605.20456, 2026. \url{https://arxiv.org/abs/2605.20456}

\bibitem{ref2}
J. Huang et al., ``Proof-or-Stop: Don't Trust the Agent, Trust the Evidence - Loop Engineering for Verifiable Evidence-Gated Lifecycle Control,'' arXiv:2607.14890, 2026. \url{https://arxiv.org/abs/2607.14890}

\bibitem{ref3}
O. Solozobov, ``Decision Evidence Maturity Model for Agentic AI: A Property-Level Method Specification,'' arXiv:2605.04093, 2026. \url{https://arxiv.org/abs/2605.04093}

\bibitem{ref4}
O. Solozobov, ``DEMM-Bench: A Cross-Regime Benchmark for Agent-Runtime Governance-Evidence Sufficiency,'' arXiv:2606.20634, 2026. \url{https://arxiv.org/abs/2606.20634}

\bibitem{ref5}
J. Cuneo, D. Chun, and G. Khanna, ``AI-GRACE: A Use-Case Operationalization Framework for Agentic AI: From Organizational Objectives and Obligations to Deployment Capabilities and Architecture,'' arXiv:2609.21192, 2026. \url{https://arxiv.org/abs/2609.21192}

\bibitem{ref6}
C. Jyu, S. Liu, and R. Jabbarvand, ``Graphectory Viewer: A Tool for Process-Centric Analysis of Agentic Software Trajectories,'' arXiv:2608.17195, 2026. \url{https://arxiv.org/abs/2608.17195}

\bibitem{ref7}
Z.-G. Xu and G. Qin, ``LLM-assisted development of Rust for high-performance bioinformatics software: practices, workflows, and boundaries,'' \emph{Genomics Communications}, vol.~3, e018, 2026. \url{https://doi.org/10.48130/gcomm-0026-0018}

\bibitem{ref8}
S. Thakur, ``Secure Policy-as-Code Guardrails for Agentic Identity Management,'' 2026. \url{https://doi.org/10.13140/RG.2.2.22830.47686}

\bibitem{ref9}
S. Thakur, ``Secure Explainable Audit Trails for Workflows in Agentic AI,'' SSRN 7046079, 2026. \url{https://papers.ssrn.com/sol3/papers.cfm?abstract_id=7046079}

\bibitem{ref10}
A. Joshi, T. Finin, K. P. Joshi, and L. Kagal, ``Deontic Policies for Runtime Governance of Agentic AI Systems,'' arXiv:2606.19464, 2026. \url{https://arxiv.org/abs/2606.19464}

\bibitem{ref11}
J. C. Davis et al., ``Cheap Code, Costly Judgment: A Case Study on Governable Agentic Software Engineering,'' arXiv:2607.01087, 2026. \url{https://arxiv.org/abs/2607.01087}

\bibitem{ref12}
S. Thakur, ``Blind Model Checking of Workflows in Agentic AI with Zero-Knowledge Proof,'' SSRN 6736917, 2026. \url{https://papers.ssrn.com/sol3/papers.cfm?abstract_id=6736917}

\bibitem{ref13}
J. He and D. Yu, ``Cognitive Admission Control: Risk-Conditioned Assurance for Consequential Actions in Agentic Distributed Systems,'' arXiv:2609.16313, 2026. \url{https://arxiv.org/abs/2609.16313}

\bibitem{ref14}
S. Yu, C. Fang, and Z. Chen, ``Engineering Agent-Integrated Software: Interaction Contracts and Continuous Assurance,'' arXiv:2609.11381, 2026. \url{https://arxiv.org/abs/2609.11381}

\bibitem{ref15}
G. Lupo, B. Q. Vo, and N. Locke, ``Trustworthy AI Posture (TAIP): A Framework for Continuous AI Assurance of Agentic Systems at Horizontal and Vertical scale,'' arXiv:2603.03340, 2026. \url{https://arxiv.org/abs/2603.03340}

\bibitem{ref16}
P. Dantas, L. Cordeiro, E. Nowroozi, and T. Norbert, ``Toward Safe LLM Agents: A Survey of Specification, Verification, and Enforcement,'' arXiv:2608.14590, 2026. \url{https://arxiv.org/abs/2608.14590}

\bibitem{ref17}
M. Kaptein, V.-J. Khan, and A. Podstavnychy, ``Runtime Governance for AI Agents: Policies on Paths,'' arXiv:2603.16586, 2026. \url{https://arxiv.org/abs/2603.16586}

\bibitem{ref18}
S. Jamshidi et al., ``Verifiable Manifest Signing and Transparency Enforcement for Secure MCP-Based LLM Pipelines,'' arXiv:2601.23132, 2026. \url{https://arxiv.org/abs/2601.23132}

\bibitem{ref19}
H. Bhati, ``Beyond Code Generation: Reliability, Verification, and Cost Economics in the Agentic Software Development Lifecycle,'' arXiv:2609.04681, 2026. \url{https://arxiv.org/abs/2609.04681}

\bibitem{ref20}
OpenHands, ``Hooks'' and OpenHands platform documentation, accessed Sep.~2026. \url{https://docs.openhands.dev/openhands/usage/customization/hooks}; X. Wang et al., ``OpenHands,'' arXiv:2407.16741.

\bibitem{ref21}
OpenAI, ``Running Codex Safely at OpenAI,'' accessed Sep.~2026. \url{https://openai.com/index/running-codex-safely/}

\bibitem{ref22}
Anthropic, ``Claude Code Security,'' accessed Sep.~2026. \url{https://code.claude.com/docs/en/security}

\bibitem{ref23}
GitHub, ``About hooks for GitHub Copilot,'' accessed Sep.~2026. \url{https://docs.github.com/en/copilot/concepts/agents/hooks}

\bibitem{ref24}
Cursor, ``Hooks,'' accessed Sep.~2026. \url{https://cursor.com/docs/hooks}

\bibitem{ref25}
Windsurf, ``Cascade Hooks,'' accessed Sep.~2026. \url{https://docs.windsurf.com/windsurf/cascade/hooks}

\bibitem{ref26}
Google, ``Jules: Reviewing plans and giving feedback,'' accessed Sep.~2026. \url{https://jules.google/docs/review-plan/}

\bibitem{ref27}
Cognition, ``Devin Advanced Capabilities,'' accessed Sep.~2026. \url{https://docs.devin.ai/work-with-devin/advanced-capabilities}

\bibitem{ref28}
Cline, ``Permission Handling,'' accessed Sep.~2026. \url{https://docs.cline.bot/sdk/guides/permission-handling}; Aider, ``Git integration,'' \url{https://aider.chat/docs/git.html}

\bibitem{ref29}
Nool, ``Architecture'' and ``Persistent Agent Memory,'' accessed Sep.~2026. \url{https://www.nool.dev/architecture}

\bibitem{ref30}
LangChain, ``LangSmith Observability,'' accessed Sep.~2026. \url{https://docs.langchain.com/langsmith/observability}

\bibitem{ref31}
Arize, ``Phoenix Overview and Tracing,'' accessed Sep.~2026. \url{https://arize.com/docs/phoenix/get-started}

\bibitem{ref32}
OpenTelemetry, ``Semantic conventions for generative AI systems,'' accessed Sep.~2026. \url{https://opentelemetry.io/docs/specs/semconv/gen-ai/}

\bibitem{ref33}
Open Policy Agent, ``Gatekeeper Operations and Admission,'' accessed Sep.~2026. \url{https://open-policy-agent.github.io/gatekeeper/website/docs/operations/}

\bibitem{ref34}
SLSA, ``Provenance v1.2,'' accessed Sep.~2026. \url{https://slsa.dev/spec/v1/provenance}

\bibitem{ref35}
in-toto, ``Attestation and software supply-chain framework,'' accessed Sep.~2026. \url{https://in-toto.io/}

\bibitem{ref36}
Sigstore, ``Bundle format and verification,'' accessed Sep.~2026. \url{https://docs.sigstore.dev/about/bundle/}; GitHub, ``Using artifact attestations,'' \url{https://docs.github.com/en/actions/security-for-github-actions/using-artifact-attestations}

\bibitem{ref37}
CycloneDX, ``Machine Learning Bill of Materials'' and ``BOM-Link,'' accessed Sep.~2026. \url{https://cyclonedx.org/capabilities/mlbom/}

\bibitem{ref38}
SPDX, ``AI and Dataset Profiles in SPDX 3,'' accessed Sep.~2026. \url{https://spdx.dev/learn/areas-of-interest/ai/}

\bibitem{ref39}
GoogleCloudPlatform, \texttt{k8s-aibom}, accessed Sep.~2026. \url{https://github.com/GoogleCloudPlatform/k8s-aibom}

\bibitem{ref40}
Model Context Protocol, ``Specification 2026-07-28,'' 2026. \url{https://modelcontextprotocol.io/specification/2026-07-28}

\bibitem{ref41}
Agent2Agent Protocol, ``Specification,'' accessed Sep.~2026. \url{https://a2a-protocol.org/latest/specification/}

\bibitem{ref42}
AG-UI, ``Agent User Interaction Protocol Specification 1.0,'' 2026. \url{https://docs.ag-ui.com/spec/1.0}

\bibitem{ref43}
NIST, ``Open Security Controls Assessment Language (OSCAL),'' accessed Sep.~2026. \url{https://pages.nist.gov/OSCAL/}

\bibitem{ref44}
W3C, ``PROV-O: The PROV Ontology,'' W3C Recommendation. \url{https://www.w3.org/TR/prov-o/}

\bibitem{ref45}
Object Management Group, ``Structured Assurance Case Metamodel (SACM),'' accessed Sep.~2026. \url{https://www.omg.org/spec/SACM/}

\bibitem{ref46}
The Assurance Case Working Group, ``Goal Structuring Notation Community Standard,'' accessed Sep.~2026. \url{https://scsc.uk/gsn}

\bibitem{ref47}
NIST, ``Artificial Intelligence Risk Management Framework,'' and ``Generative AI Profile,'' accessed Sep.~2026. \url{https://www.nist.gov/itl/ai-risk-management-framework}

\bibitem{ref48}
European Union, Regulation (EU) 2024/1689, consolidated text 27 Jul.~2026; OWASP, ``Top 10 for Agentic Applications 2026.'' \url{https://eur-lex.europa.eu/legal-content/EN/TXT/HTML/?uri=CELEX:02024R1689-20260727}; \url{https://genai.owasp.org/resource/owasp-top-10-for-agentic-applications-for-2026/}

\bibitem{ref49}
P. Zheng et al., ``SWE-Bench Pro Verified: A Reliable Benchmark for Software Engineering Agents,'' arXiv:2609.08149, 2026. \url{https://arxiv.org/abs/2609.08149}

\bibitem{ref50}
E. Debenedetti et al., ``AgentDojo: A Dynamic Environment to Evaluate Prompt Injection Attacks and Defenses for LLM Agents,'' arXiv:2406.13352, 2024. \url{https://arxiv.org/abs/2406.13352}

\bibitem{ref51}
Y. Ruan et al., ``Identifying the Risks of LM Agents with an LM-Emulated Sandbox (ToolEmu),'' arXiv:2309.15817, 2024. \url{https://arxiv.org/abs/2309.15817}

\bibitem{ref52}
F. Alpay and T. Alpay, ``AgentSecBench: Measuring Prompt Injection, Privacy Leakage, and Tool-Use Integrity in LLM Agents,'' arXiv:2605.26269, 2026. \url{https://arxiv.org/abs/2605.26269}

\bibitem{ref53}
L. Ba, Q. Li, and S. Li, ``CIBER: A Comprehensive Benchmark for Security Evaluation of Code Interpreter Agents,'' arXiv:2602.19547, 2026. \url{https://arxiv.org/abs/2602.19547}

\bibitem{ref54}
T. King, S. Kehrberg, M. Beigl, and T. R\"{o}ddiger, ``pcbGPT: Automatic PCB Schematic Synthesis from Natural Language Requirements,'' arXiv:2606.01188, 2026. \url{https://arxiv.org/abs/2606.01188}

\bibitem{ref55}
Agile-V, \texttt{agile\_v\_skills}, repository snapshot \texttt{982c1f6435219f0fe83dcef485503f45a7e9717e}, and \texttt{agentic\_agile\_v}, repository snapshot \texttt{54eaa5451fcbfe9fa1c59355584deddbf7cef342}, accessed 23 Sep.~2026. \url{https://github.com/Agile-V/agile_v_skills}; \url{https://github.com/Agile-V/agentic_agile_v}

\bibitem{ref56}
GitLab, ``GitLab Duo Agent Platform'' and ``AI Governance,'' accessed Sep.~2026. \url{https://docs.gitlab.com/user/duo_agent_platform/}; \url{https://docs.gitlab.com/user/ai-governance/}

\bibitem{ref57}
SonarSource, ``AI Code Assurance'' and ``Quality gates for AI code,'' accessed Sep.~2026. \url{https://docs.sonarsource.com/sonarqube-server/project-administration/ai-features/set-up-ai-code-assurance}; \url{https://docs.sonarsource.com/sonarqube-server/2025.3/quality-standards-administration/ai-code-assurance/quality-gates-for-ai-code}

\bibitem{ref58}
Qodo, ``Custom Compliance,'' ``Rule System,'' and ``Agentic PR Review,'' accessed Sep.~2026. \url{https://docs.qodo.ai/qodo-documentation/code-review/qodo-merge/features/custom-compliance}; \url{https://docs.qodo.ai/qodo-release-notes/release-notes/qodo-ide-plugin-latest-release}

\bibitem{ref59}
Semgrep, ``Semgrep Guardian'' and ``Semgrep Workflows,'' accessed Sep.~2026. \url{https://semgrep.dev/products/semgrep-guardian/}; \url{https://semgrep.dev/products/semgrep-workflows}

\bibitem{ref60}
Snyk, ``Developer Guardrails for Agentic Workflows'' and ``Snyk Studio Usage Analytics,'' accessed Sep.~2026. \url{https://docs.snyk.io/integrations/developer-guardrails-for-agentic-workflows}; \url{https://docs.snyk.io/integrations/snyk-studio-agentic-integrations/usage-analytics}

\bibitem{ref61}
Tabnine, ``Tabnine Agent,'' ``Tabnine CLI,'' and enterprise context documentation, accessed Sep.~2026. \url{https://docs.tabnine.com/main/getting-started/tabnine-agent}; \url{https://docs.tabnine.com/main/getting-started/tabnine-cli}
\end{thebibliography}
\end{document}